\documentclass[pra,twocolumn,superscriptaddress,preprintnumbers,showpacs,byrevtex,floatfix]{revtex4}
\usepackage{graphicx,amssymb,amsmath}
\usepackage{amsmath}
\usepackage{amsfonts}
\usepackage{amssymb}
\usepackage{graphicx}
\def\barhex{16}
\def\kbar
  {{\mathchoice
      {\hbox{\lower.07em \hbox{$\mathchar"\barhex$}}}
      {\hbox{\lower.07em \hbox{$\mathchar"\barhex$}}}
      {\hbox{\lower.049em\hbox{$\scriptstyle\mathchar"\barhex$}}}
      {\hbox{\lower.035em\hbox{$\scriptscriptstyle\mathchar"\barhex$}}}%
  \mkern-8muk}}
\begin{document}

\title{Analysis of dynamical tunnelling experiments with a Bose-Einstein condensate }
\author{W. K. Hensinger}
\email{hensinge@umich.edu}\altaffiliation[Present address:
]{Department of Physics, University of Michigan, 2477 Randall
Laboratory, 500 East University Ave., Ann Arbor, MI 48109-1120,
USA}\affiliation{Centre for Biophotonics and Laser Science,
Department of Physics, The University of Queensland, Brisbane,
Queensland 4072, Australia}
 \affiliation{National Institute of
Standards and Technology, Gaithersburg, Maryland, 20899, USA}
\author{A. Mouchet}\email{mouchet@celfi.phys.univ-tours.fr}
\affiliation{Laboratoire de Math\'{e}matique et de Physique
Th\'{e}orique (CNRS UMR 6083), Avenue Monge, Parc de Grandmont,
37200 Tours, France}
\author{P. S. Julienne}
\affiliation{National Institute of Standards and Technology,
Gaithersburg, Maryland, 20899, USA}
\author{D. Delande}
\affiliation{Laboratoire Kastler-Brossel (CNRS UMR 8552),
Universit\'{e} Pierre et Marie Curie, 4 place Jussieu, F-75005
Paris, France}
\author{N. R. Heckenberg}
\affiliation{Centre for Biophotonics and Laser Science, Department
of Physics, The University of Queensland, Brisbane, Queensland
4072, Australia}
\author{H. Rubinsztein-Dunlop}
\affiliation{Centre for Biophotonics and Laser Science, Department
of Physics, The University of Queensland, Brisbane, Queensland
4072, Australia} \email{hensinge@physics.uq.edu.au}
\begin{abstract}
Dynamical tunnelling is a quantum phenomenon where a classically
forbidden process occurs, that is prohibited not by energy but by
another constant of motion. The phenomenon of dynamical tunnelling
has been recently observed in a sodium Bose-Einstein condensate.
We present a detailed analysis of these experiments using
numerical solutions of the three dimensional Gross-Pitaevskii
equation and the corresponding Floquet theory. We explore the
parameter dependency of the tunnelling oscillations and we move
the quantum system towards the classical limit in the
experimentally accessible regime.
\end{abstract}

 \maketitle

\section{Introduction}

Cold atoms provide a system which is particularly suited to study
quantum nonlinear dynamics, quantum chaos and the
quantum-classical borderland. On relevant timescales the effects
of decoherence and dissipation are negligible. This allows us to
study a Hamiltonian quantum system.  Only recently dynamical
tunnelling was observed in experiments with ultra-cold atoms
\cite{Hensinger2001c,Hensinger03}. ``Conventional'' quantum
tunnelling allows a particle to pass through a classical energy
barrier.  In contrast, in dynamical tunnelling a constant of
motion other than energy classically forbids to access a different
motional state. In our experiments atoms tunnelled back and forth
between their initial oscillatory motion and the motion 180$
{{}^\circ}$ out of phase. A related experiment was carried out by
Steck, Oskay and Raizen \cite{Steck2001,Steck2002} in which atoms
tunnelled from one uni-directional librational motion into another
oppositely directed motion.

Luter and Reichl \cite{Luter02} analyzed both experiments
calculating mean momentum expectations values and Floquet states
for some of the parameter sets for which experiments were carried
out and found good agreement with the observed tunnelling
frequencies. Averbukh, Osovski and Moiseyev \cite{Averbukh02}
pointed out that it is possible to effectively control the
tunnelling period by varying the effective Planck's constant by
only 10\%. They showed one can observe both suppression due to the
degeneracy of two Floquet states and enhancement due to the
interaction with a third state in such a small interval.

Here we present a detailed theoretical and numerical analysis of
our experiments. We use numerical solutions of the
Gross-Pitaevskii equation and Floquet theory to analyze the
experiments and to investigate the relevant tunnelling dynamics.
In particular we show how dynamical tunnelling can be understood
in a two and three state framework using Floquet theory.  We show
that there is good agreement between experiments and both
Gross-Pitaevskii evolution and Floquet theory. We examine the
parameter sensitivity of the tunnelling period to understand the
underlying tunnelling mechanisms. We also discuss such concepts as
chaos-assisted and resonance-assisted tunnelling in relation to
our experimental results. Finally predictions are made concerning
what can happen when the quantum system is moved towards the
classical limit.

In our experiments a sodium Bose-Einstein condensate was
adiabatically loaded into a far detuned optical standing wave.
 For a sufficient
large detuning, spontaneous emission can be neglected on the time
scales of the experiments (160~$\mu$s). This also allows to
consider the external degrees of freedom only. The dynamics
perpendicular to the standing wave are not significant, therefore
we are led to an effectively one-dimensional system. The
one-dimensional system can be described in the corresponding
two-dimensional phase space which is spanned by momentum and
position coordinates along the standing wave. Single frequency
modulation of the intensity of the standing wave leads to an
effective Hamiltonian for the center-of-mass motion given by
\begin{equation}
H=\frac{p_{x}^{2}}{2m}+\frac{\hbar\Omega_{\mathrm{eff}}}{4}\left(
{1-2\varepsilon\sin}\left(  {\omega t+\phi}\right)  \right)  \sin
^{2}(kx)\label{firstH}%
\end{equation}
where the effective Rabi frequency is $\Omega_{\mathrm{eff}}=\Omega^{2}%
/\delta$, $\Omega=\Gamma\sqrt{I/I_{\mathrm{sat}}}$ is the resonant
Rabi frequency, $\varepsilon$ is the modulation parameter,
$\omega$ is the modulation angular frequency, $\Gamma$ is the
inverse spontaneous lifetime, $\delta$ is the detuning of the
standing wave, $t$ is the time, $p_{x}$ the momentum component of
the atom along the standing wave, and $k$ is the wave number. Here
$I$ is the spatial-mean of the intensity of the unmodulated
standing wave (which is half of the peak intensity) so
$\Omega=\Gamma
\sqrt{I_{\mathrm{peak}}/2I_{\mathrm{sat}}}$ where $I_{\mathrm{sat}}%
=hc\Gamma/\lambda^{3}$ is the saturation intensity. $\lambda$ is
the wavelength of the standing wave. $\phi$ determines the start
phase of the amplitude modulation. Using scaled variables
\cite{Moore95} the Hamiltonian is given by
\begin{equation}
\mathcal{H}=p^{2}/2+2\kappa(1-2\varepsilon\sin\left(
\tau+\phi\right)
)\sin^{2}(q/2)\label{scaled}%
\end{equation}
where $\mathcal{H}=(4k^{2}/m\omega^{2})H$, $q=2kx$, and
$p=(2k/m\omega)p_{x}$.

\begin{figure}[ptbh]
\centering
\includegraphics[width=5cm,keepaspectratio]{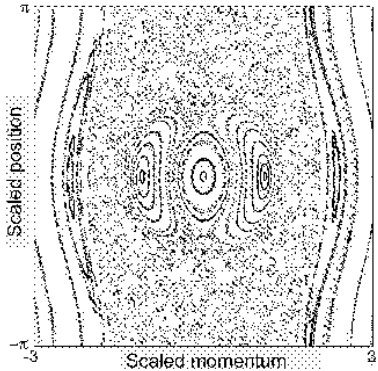}%
\caption{Poincar\'{e} section for a classical particle in an
amplitude-modulated optical standing wave. Momentum and position
(one well of the standing wave) of the particle along the standing
wave are plotted stroboscopically with the stroboscopic period
being equal to the modulation period. The central region consists
of small amplitude motion. Chaos (dotted region) separates this
region from two period-1 regions of regular motion (represented in
the Poincar\'{e} section as sets of closed curves) located left
and right of the centre along momentum $p=0$ . Further out in
momentum are two stable regions of motion known as librations. At
the edges are bands of regular motion corresponding to above
barrier motion. It is plotted for modulation parameter
$\varepsilon=0.20$ and scaled well depth $\kappa=1.20$.}%
\label{poincare}%
\end{figure}

The driving amplitude is given by
\begin{equation}
\kappa=\omega_{r}\Omega_{\mathrm{eff}}/\omega^{2}=\frac{\hbar k^{2}%
\Omega_{\mathrm{eff}}}{2\omega^{2}m}=\frac{4U_{0}\omega_{r}
}
{\omega^{2}\hbar}
\end{equation}
where $\omega_{r}=\hbar k^{2}/2m$ is the recoil frequency,
$\tau=t\omega$ is the scaled time variable and $U_{0}$ is the well
depth. The commutator of scaled position and momentum is given by
\begin{equation}
\left[  p,q\right]  =i\kbar,
\end{equation}
where the scaled Planck's constant is
$\kbar=8\omega_{r}/\omega.$
 For $\kappa=1.2$ and
$\varepsilon=0.20$ the classical Poincar\'e surface of section is
shown in Fig.~\ref{poincare}.
Two symmetric regular regions can be observed about~$(q=0,p=1)$
and~$(q=0,p=-1)$. These regions correspond to oscillatory motion
in phase with the amplitude modulation in each well of the
standing wave.  In the experiment
\cite{Hensinger2001c,Hensinger03} atoms are loaded in a period-1
region of regular motion by controlling their initial position and
momentum and by choosing the starting phase of the amplitude
modulation appropriately.  Classically atoms should retain their
momentum state when observed stroboscopically (time step is one
modulation period). A distinct signature of dynamical tunnelling
is a coherent oscillation of the stroboscopically observed mean
momentum as shown in Fig. \ref{chap12fig8} and reported in ref.
\cite{Hensinger2001c}.

In Sec. \ref{one} we introduce the theoretical tools to analyze
dynamical tunnelling by discussing Gross-Pitaevskii simulations
and the appropriate Floquet theory. We present a thorough analysis
of the experiments from ref. \cite{Hensinger2001c} in Sec.
\ref{two}. After showing some theoretical results for the
experimental parameters we give a small overview of what to expect
when some of the system parameters in the experiments are varied
in Sec. \ref{three} and give some initial analysis. In Sec.
\ref{four} we point to pathways to analyze the quantum-classical
transition for our experimental system and give conclusions in
Sec. \ref{five}.

\section{Theoretical analysis of the dynamic evolution of a Bose-Einstein
condensate}\label{one}

\subsection{Dynamics using the Gross-Pitaevskii equation}

The dynamics of a Bose-Einstein condensate in a time-dependent
potential in the mean-field limit are described by the
Gross-Pitaevskii equation \cite{Dalfovo99,Parkins97}

\begin{align}
i\hbar\frac{\partial\Psi\left(  \mathbf{r},t\right)  }{\partial t}  & =\bigg
[-\frac{\hbar}{2m}\nabla^{2}+V_{trap}\left(  \mathbf{r},t\right)  +V\left(
\mathbf{r},t\right) \nonumber\\
& \hspace{1cm}+N\frac{4\pi\hbar^{2}a}{m}\left|  \Psi\left(  \mathbf{r}%
,t\right)  \right|  ^{2}\bigg]\Psi\left(  \mathbf{r},t\right)
\label{Grossp}\end{align}
\newline where $N$ is the mean number of atoms in the condensate, $a$ is the
scattering length with $a=2.8$ nm for sodium.
 $V_{trap}\left(  \mathbf{r}\right)  $ is the
trapping potential which is turned off during the interaction with
the standing wave and $V\left( \mathbf{r},t\right)$ is the time
dependent optical potential induced by the optical standing wave.
The Gross-Pitaevskii equation is propagated in time using a
standard numerical split-operator, fast Fourier transform method.
The size of the spatial grid of the numerical simulation is chosen
to contain the full spatial extent of the initial condensate
(therefore all the populated wells of the standing wave), and the
grid has periodic boundary conditions at each side (a few
unpopulated wells are also included on each side).

To obtain the initial wave function a Gaussian test function is
evolved by imaginary time evolution to converge to the ground
state of the stationary Gross-Pitaevskii equation. Then the
standing wave is turned on adiabatically with $V\left(
\mathbf{r},t\right) $ approximately having the form of a linear
ramp. After the adiabatic turn-on, the condensate wave function is
found to be localized at the bottom of each well of the standing
wave. The standing wave is shifted and the time-dependent
potential now has the form
\begin{equation}
V\left(  \mathbf{r},t\right)
=\frac{\hbar\Omega_{\mathrm{eff}}}{4}\left(
{1-2\varepsilon\sin}\left(  {\omega t+\phi}\right)  \right)
\sin^{2}(kx+\varphi)
\end{equation}
where $\varphi$ is the phase shift which is applied to selectively load one
region of regular motion \cite{Hensinger2001c}. The position representation of
the atomic wave function $\left|  \Psi\left(  \mathbf{r}\right)  \right|
^{2}$ just before the modulation starts (and after the phase shift) is shown
in Fig. \ref{chap12fig22} ($t=0T$).

The Gross-Pitaevskii equation is used to model the experimental
details of a Bose-Einstein condensate in an optical 1-D lattice.
The experiment effectively consists of many coherent single-atom
experiments. The coherence is reflected in the occurrence of
diffraction peaks in the atomic momentum distribution (see Fig.
\ref{chap12fig24}). Utilizing the Gross-Pitaevskii equation, the
interaction between these single-atom experiments is modelled by a
classical field. Although ignoring quantum phase fluctuations in
the condensate, the wave nature of atoms is still contained in the
Gross-Pitaevskii equation and dynamical tunnelling is a quantum
effect that results from the wave nature of the atoms. The
assumption for a common phase for the whole condensate is well
justified for the experimental conditions as the timescales of the
experiment and the lattice well depth are sufficiently small. It
will be shown in the following (see Fig.~\ref{chap12fig25}) that
the kinetic energy is typically of the order of~$10^{5}$ Hz which
is much larger than the non-linear term in the Gross-Pitaevskii
equation \ref{Grossp} which is on the order of~400 Hz. The
experimental results, in particular dynamical tunnelling could
therefore be modelled by a single particle Schr\"{o}dinger
equation in a one-dimensional single well with periodic boundary
conditions. Nevertheless the Gross-Pitaevskii equation is used to
model all the experimental details of a Bose-Einstein condensate
in an optical lattice to guarantee maximum accuracy. We will
discuss and compare the Gross-Pitaevskii and the Floquet
approaches below (last paragraph of section~II).

Theoretical analysis of the dynamical tunnelling experiments will
be presented in this paper utilizing numerical solutions of the
Gross-Pitaevskii equation.  Furthermore we will analyze the system
parameter space which is spanned by the scaled well depth
$\kappa,$ the modulation parameter $\varepsilon$ and the scaled
Planck's constant $\kbar$. In fact variation of the scaled
Planck's constant in the simulations allows one to move the
quantum system towards the classical limit.

\subsection{Floquet analysis\label{Floquet}}

The quantum dynamics of a periodically driven Hamiltonian system
can be described in terms of the eigenstates of the Floquet
operator $F$, which evolves the system in time by one modulation
period. In the semiclassical regime, the Floquet eigenstates can
be associated with regions of regular and irregular motion of the
classical map. However when $\hbar$ is not sufficiently small
compared to a typical classical action the phase-space
representation of the Floquet eigenstates do not necessarily match
with some classical (regular or irregular)
structures\cite{Dyrting93, Mouchet2001}. However, initial states
localized at the stable region around a fixed point in the
Poincar\'{e} section can be associated with superpositions of a
small number of Floquet eigenstates. Using this state basis, one
can reveal the analogy of the dynamical tunnelling experiments and
conventional tunnelling in a double well system. Two states of
opposite parity which can be responsible for the observed
dynamical tunnelling phenomenon are identified.  Floquet states
are stroboscopic eigenstates of the system. Their phase space
representation therefore provides a quantum analogue to the
classical stroboscopic phase space representation, the
Poincar\'{e} map.

Only very few states are needed to describe the evolution of a
wave packet that is initially strongly localized on a region of
regular motion. A strongly localized wavepacket is used in the
experiments (strongly localized in each well of the standing wave)
making the Floquet basis very useful. In contrast, describing the
dynamics in momentum or position representation requires a large
number of states so that some of the intuitive understanding which
one can obtain in the Floquet basis is impossible to gain. For
example, the tunnelling period can be derived from the
quasi-eigenenergies of the relevant Floquet states, as will be
shown below.

For an appropriate choice of parameters the phase space exhibits two
period-1 fixed points, which for a suitable Poincar\'{e} section lie
on the momentum axis at $\pm p_{0}$, as in reference
\cite{Hensinger2001c}. For certain values of the scaled well depth
$\kappa$ and $\ $modulation parameter $\varepsilon$ there are two
dominant Floquet states $|\phi_{\pm}\rangle$ that are localized on
both fixed points but are distinguished by being even or odd
eigenstates of the parity operator that changes the sign of
momentum. A state localized on just one fixed point is therefore
likely to have dominant support on an even or odd superposition of
these two Floquet states:
\begin{equation}
|\psi\left(  \pm p_{0}\right)  \rangle\approx\left(  |\phi_{+}\rangle\pm
|\phi_{-}\rangle\right)  /\sqrt{2}.
\end{equation}

The stroboscopic evolution is described by repeated application of the
Floquet operator. As this is a unitary operator,
\begin{equation}
F|\phi_{\pm}\rangle=e^{\left(  -i2\pi\phi_{\pm}/\kbar
\right)  }|\phi_{\pm}\rangle.
\end{equation}
$\phi_\pm$ are the Floquet quasienergies.  Thus at a time which is
$n$ times the period of modulation, the state initially localized
on $+p_{0}$, evolves to
\begin{equation}
|\psi\left(  n\right)  \rangle\approx\left(  e^{\left(  -i2\pi n\phi
_{+}/
\kbar
\right)  }|\phi_{+}%
\rangle+e^{\left(  -i2\pi n\phi_{-}/
\kbar
\right)  }|\phi_{-}\rangle\right)  /\sqrt{2}.
\end{equation}
Ignoring an overall phase and defining the separation between
Floquet quasienergies as
\begin{equation}
\label{eq:splitting}
\Delta\phi=\phi_{-}-\phi_{+},
\end{equation}
one obtains
\begin{equation}
|\psi\left(  n\right)  \rangle\approx\left(  |\phi_{+}\rangle+e^{\left(
-i2\pi n\Delta\phi/
\kbar
\right)
}|\phi_{-}\rangle\right)  /\sqrt{2}.
\end{equation}
At
\begin{equation}
n=\kbar/2\Delta\phi\label{period}%
\end{equation}
periods, the state will form the anti-symmetric superposition of
Floquet states and thus is localized on the other fixed point at
$-p_{0}$. In other words the atoms have tunnelled from one of the
fixed points to the other. This is reminiscent of barrier
tunnelling between two wells, where a particle in one well, in a
superposition of symmetric and antisymmetric energy eigenstates,
oscillates between wells with a frequency given by the energy
difference between the eigenstates.

Tunnelling can also occur when the initial state has significant
overlap with two non-symmetric states. For example, if the initial
state is localized on two Floquet states, one localized inside the
classical chaotic region and one inside the region of regular
motion, a distinct oscillation in the stroboscopic evolution of
the mean momentum may be visible. The frequency of this tunnelling
oscillation depends on the spacing of the corresponding
quasi-eigenenergies in the Floquet spectrum. In many cases
multiple tunnelling frequencies occur in the stroboscopic
evolution of the mean momentum some of which are due to tunnelling
between non-symmetric states.

Quantum dynamical tunnelling may be defined in that a particle can
access a region of phase space in a way that is forbidden by the
classical dynamics.  This implies that it crosses a
Kolmogorov-Arnold-Moser (KAM) surface \cite{Berry78a,Arnold79}.
The clearest evidence of dynamical tunnelling can be obtained by
choosing the scaled Planck's constant $\kbar$ sufficiently small
so that the atomic wavefunction is much smaller than the region of
regular motion (the size of the wave function is given by
$\kbar$). Furthermore it should be centered inside the region of
regular motion.  However, even if the wave packet is larger than
the region of regular motion and also populates the classical
chaotic region of phase space one can still analyze
quantum-classical correspondence and tunnelling. One assumes a
classical probability distribution of point particles with the
same size as the quantum wave function and compares the classical
evolution of this point particle probability distribution with the
quantum evolution of the wave packet. A distinct difference
between the two evolutions results from the occurrence of
tunnelling assuming that the quantum evolution penetrates a KAM
surface visibly.

\subsection{Comparison}

Using the Gross-Pitaevskii equation one can exactly simulate the
experiment because the momentum or position representation is
used. Therefore the theoretical simulation can be directly
compared with the experimental result. In contrast Floquet states
do not have a straightforward experimental intuitive analogue. In
the Floquet states analysis one can compare the quasienergy
splitting between the tunnelling Floquet state with the
experimentally measured tunnelling period. The occurrence of
multiple frequencies in the experimentally observed tunnelling
oscillations might also be explained with the presence of more
than two dominant Floquet states, the tunnelling frequencies being
the energy splitting between different participating Floquet
states. Using the Gross-Pitaevskii approach one can simulate the
experiment with high precision. The same number of populated wells
as in the experiment can be used and the turn-on of the standing
wave can be simulated using an appropriate turn-on Hamiltonian.
Therefore one can obtain the correct initial state with high
precision. Furthermore the mean-field interaction is simulated
which is not contained in the Floquet approach.

\section{Numerical simulations of the experiments}\label{two}

\subsection{Gross-Pitaevskii simulations}

In this section experimental results that were discussed in a previous
paper \cite{Hensinger2001c} are compared with numerical simulations of
the Gross-Pitaevskii (GP) equation.

\begin{figure}[ptbh]
\centering
\includegraphics[width=7cm,keepaspectratio]{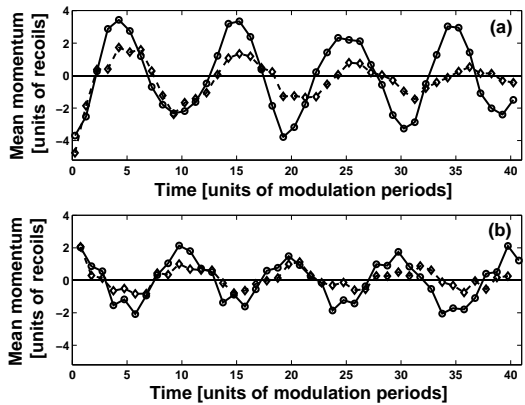}%
\caption{Stroboscopic mean momentum as a function of the interaction
time with the modulated standing wave measured in modulation periods,
$n$, for modulation parameter $\varepsilon=0.29$, scaled well depth
$\kappa=1.66$, modulation frequency $\omega/2\pi=250$ kHz and a phase
shift $\varphi=0.21\cdot2\pi.$ (a) and (b) correspond to the two
different interaction times $n+0.25$ and $n+0.75$ modulation periods,
respectively. Results from the dynamic evolution of the
Gross-Pitaevskii equation are plotted as solid line (circles) and the
experimental data are plotted as dashed line (diamonds).}%
\label{chap12fig8}%
\end{figure}

In the experiment \cite{Hensinger2001c} the atomic wave function was prepared
initially to be localized around a period-1 region of regular motion. Figure
\ref{chap12fig8} shows the stroboscopically measured mean momentum as a
function of the interaction time with the standing wave for modulation
parameter $\varepsilon=0.29$, scaled well depth $\kappa=1.66$, modulation
frequency $\omega/2\pi=250$ kHz and a phase shift of $\varphi=0.21\cdot2\pi.$
 (a) and (b) correspond to the two
different interaction times $n+0.25$ (standing wave modulation
ends at a maximum of the amplitude modulation) and $n+0.75$
modulation periods (standing wave modulation ends at a minimum of
the amplitude modulation), respectively. Results from the
simulations (solid line, circles) are compared with the
experimental data (dashed line, diamonds). Dynamical tunnelling
manifests itself as a coherent oscillation of the stroboscopically
observed mean momentum. This occurs in contrast to the classical
prediction in which atoms should retain their momentum state when
observed stroboscopically (time step is one modulation period).
There is good agreement between experiment and theory as far as
the tunnelling period is concerned. However, the experimentally
measured tunnelling amplitude is smaller than the theoretical
prediction.

It should be noted that the theoretical simulations do not take
account of any possible spatial and temporal variations of the
scaled well depth (eg. light intensity), which could possibly lead
to the observed discrepancy. It was decided to produce simulations
without using any free parameters. However the uncertainty
associated with the experiment would allow small variations of the
modulation parameter $\varepsilon$ and the scaled well depth
$\kappa.$ There was a 10\% uncertainty in the value of the scaled
well depth $\kappa$ and a 5\% uncertainty in the modulation
parameter $\varepsilon$ (all reported uncertainties are 1 s.d.
combined systematic and statistical uncertainties). Both temporal
and spatial uncertainty during one run of the experiment are
contained in these values as well as the systematic total
measurement uncertainty. It was verified that there are no
important qualitative changes when varying the parameters in the
uncertainty regime for the simulations presented here. However,
the agreement between experiment and theory often can be optimized
(not always non-ambiguously, meaning that it is sometimes hard to
decide which set of parameters produces the best fit). Although
the theoretical simulation presented does not show any decay in
the mean momentum curve, a slight change of parameters inside the
experimental uncertainty can lead to decay which is most likely
caused by another dominant Floquet state whose presence leads to
the occurrence of a beating of the tunnelling oscillations which
appears as decay (and revival on longer time scales). A detailed
analysis of the corresponding Floquet states and their meaning
will be presented in Sec. \ref{Floquetanal}. An example of how a
change of parameters inside the experimental uncertainty in the
simulations can optimize the agreement between theory and
experiment is shown in Fig. \ref{chap12fig9} and Fig.
\ref{chap12fig10}. Both figures are described in more detail later
in this section. Another reason for the observed discrepancy could
be the evolution of non-condensed atoms that is not contained in
the GP approach and the interaction of non-condensed atoms with
the condensate. With a sufficiently long adiabatic turn-on time of
the far detuned standing wave the production of non-condensed
atoms should be negligible. The interaction of the condensate with
non-condensed atoms should also be negligible due to the low
atomic density (note that in the experiments the condensate is
expanded before the standing wave is turned on). However, further
studies are needed to give an exact estimate of these effects.

\begin{figure}[ptbh]
\centering
\includegraphics[width=8.5cm,keepaspectratio]{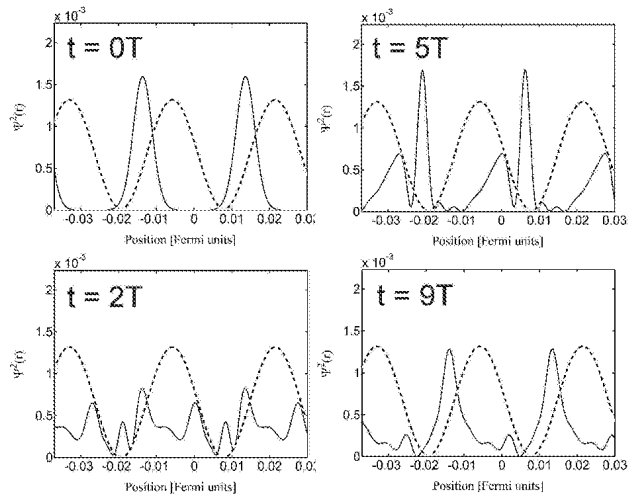}%
\caption{Position representation of the atomic wave function as a
function of the number of modulation periods calculated using the
Gross-Pitaevskii equation. The wave packet is plotted
stroboscopically at a phase so it is located classically
approximately at its highest point in the potential well. The
position of the standing wave wells is also shown (dotted line).
Dynamical tunnelling can be observed. At $t = 5T$ most atoms have
tunnelled into the other period-1 region of regular motion. The
position axis is given in Fermi units (scaled by
the mean Thomas-Fermi diameter).}%
\label{chap12fig22}%
\end{figure}

The position representation of the atomic wave function $\left|
\Psi\left( x\right)  \right|  ^{2}$ is plotted stroboscopically
after multiples of one modulation period in Fig. \ref{chap12fig22}
for the same parameters as Fig. \ref{chap12fig8}.
The position of the standing wave wells is also shown (dotted
line), their amplitude is given in arbitrary units. The position
axis is scaled with the mean Thomas-Fermi diameter. The initial
modulation phase is chosen so that the wavepacket should be
located classically approximately at its highest point of the
potential well \cite{Hensinger03}. Choosing this stroboscopic
phase the two regions of regular motion are always maximally
separated in position space. Using this phase for the stroboscopic
plots enables the observation of dynamical tunnelling in position
space as the two regions of regular motion are located to the left
and to the right of the minimum of the potential well, being
maximally spatially separated. In contrast, in the experiments,
tunnelling is always observed in momentum space (the standing wave
is turned off when the regions of regular motion are at the bottom
of the well, overlapping spatially but having oppositely directed
momenta) as it is difficult to optically resolve individual wells
of the standing wave. The first picture in Fig. \ref{chap12fig22}
($t=0T$) shows the initial wave packet before the modulation is
turned on.  Subsequent pictures exemplify the dynamical tunnelling
process. At $t=2T,$ half the atoms have tunnelled; most of the
atoms are in the other region of regular motion at $t=5T$. The
atoms have returned to their initial position at about $t=9T$. The
double peak structure at $t=5T$ and the small central peak at
$t=2T$ could indicate that Floquet states other than the two
dominant ones are also loaded which is likely as a relatively
large Planck's constant was used enabling the initial wave packet
to cover a substantial phase space area.

\begin{figure}[ptbh]
\centering
\includegraphics[width=8.5cm,keepaspectratio]{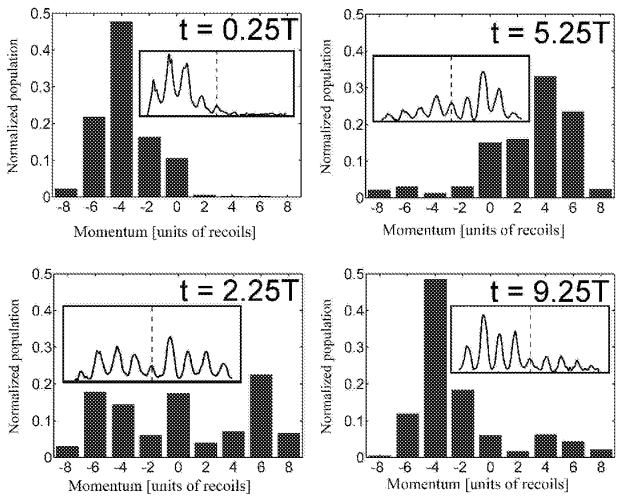}%
\caption{Momentum distributions as a function of the interaction time with the
modulated standing wave calculated using numerical solutions of the
Gross-Pitaevskii equation for the same parameters as Fig. \ref{chap12fig22}.
Initially atoms have mostly negative momentum ($t=0.25T$). After approximately
5 modulation periods most atoms populate a state with positive momentum,
therefore having undergone dynamical tunnelling. Corresponding experimental
data from reference \cite{Hensinger2001c} is also included as insets.}%
\label{chap12fig24}%
\end{figure}

Figure \ref{chap12fig24} shows simulations of the stroboscopically
measured momentum distributions $\left|  \Psi\left(  p\right)
\right|  ^{2}$ as bar graphs for the same parameters as Fig.
\ref{chap12fig22}.
Corresponding experimental data from reference
\cite{Hensinger2001c} are also included as insets. The momentum
distributions are plotted at $n+0.25$ modulation periods, where
$n$ is an integer. At this modulation phase the amplitude
modulation is at its maximum and atoms in a period-1 region of
regular motion are classically at the bottom of the well having
maximum momentum. The two period-1 regions of regular motion can
be distinguished in their momentum representation at this phase as
they have opposite momenta. At $t=0.25T$ the atoms which were
located initially half way up the potential well as shown in Fig.
\ref{chap12fig22} at $t=0T$ have ``rolled'' down the well having
acquired negative momentum. Momentum distributions for subsequent
times illustrate the dynamical tunnelling process. At $t=5.25$
modulation periods most atoms have reversed their momentum and
they return to their initial momentum state at approximately
$t=9.25T$. The simulations are in reasonable agreement with the
experimentally measured data.

\begin{figure}[ptbh] \centering
\includegraphics[width=7cm,keepaspectratio]{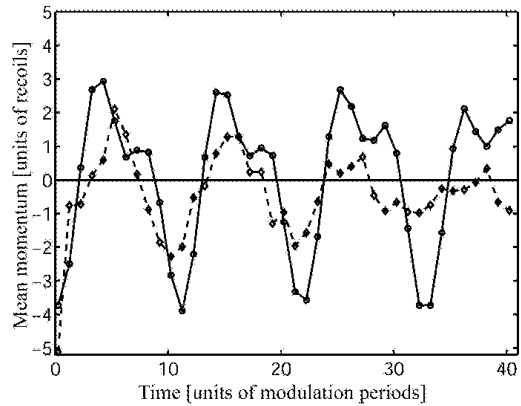}\caption{Mean
momentum as a function of the interaction time with the modulated
standing wave measured in modulation periods, $n,$ for
$\varepsilon=0.28$, $\kappa=1.49$, $\omega/2\pi=250$ kHz, and
$\varphi=0.22\cdot2\pi.$ The points are plotted stroboscopically
with an interaction time of $n+0.25$ modulation periods which
corresponds to turning off the standing wave at maximum. Results
from the dynamic evolution of the Gross-Pitaevskii equation are
plotted as solid line (circles) and the experimental data is
plotted as dashed line (diamonds).}%
\label{chap12fig15}%
\end{figure}

\begin{figure}[ptbh] \centering
\includegraphics[width=7cm,keepaspectratio]{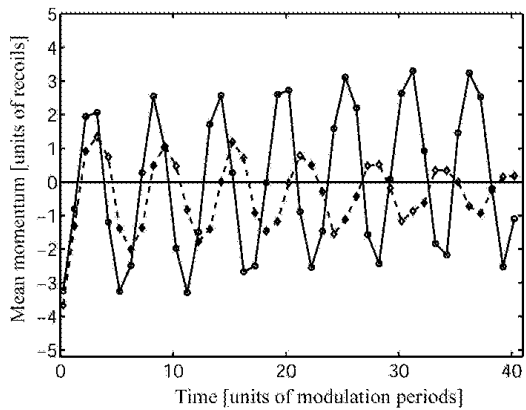}\caption{Mean
momentum as a function of the interaction time with the modulated standing
wave measured in modulation periods, $n$, for modulation parameter
$\varepsilon=0.30$, scaled well depth $\kappa=1.82$, modulation frequency
$\omega/2\pi=222$ kHz and phase shift $\varphi=0.21\cdot2\pi$. The points are
plotted stroboscopically with an interaction time of $n+0.25$ modulation
periods which corresponds to turning off the standing wave at maximum. Results
from the dynamic evolution of the Gross-Pitaevskii equation are plotted as
solid line (circles) and the experimental data is plotted as dashed line
(diamonds).}%
\label{chap12fig9}%
\end{figure}

\begin{figure}[ptbh] \centering
\includegraphics[width=7cm,keepaspectratio]{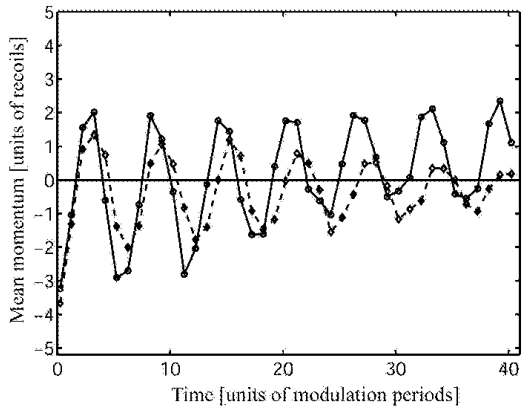}\caption{Mean
momentum as a function of the interaction time with the modulated standing
wave measured in modulation periods, $n$, using the same parameters as Fig.
\ref{chap12fig9} but the modulation parameter $\varepsilon$ is reduced from
0.30 to 0.28 for the theoretical simulation. A much better fit is obtained.}%
\label{chap12fig10}%
\end{figure}

Dynamical tunnelling is sensitive to the modulation parameter, the
scaled well depth and the scaled Planck's constant. To illustrate
this, tunnelling data along with the appropriate evolution of the
Gross Pitaevskii equation will be shown for another two parameter
sets. Even though this represents only a small overview of the
parameter dependency of the tunnelling oscillations it may help to
appreciate the variety of features in the atomic dynamics. Figure
\ref{chap12fig15} shows the theoretical simulation and the
experimental data for $\varepsilon=0.28$, $\kappa=1.49$,
$\omega/2\pi=250$ kHz, and $\varphi=0.22\cdot2\pi.$ The mean
momentum is plotted stroboscopically with the intensity modulation
at maximum ($n+0.25$ modulation periods, $n$ being an integer).
The solid line (circles) is produced by a Gross-Pitaevskii
simulation and the dashed line (diamonds) consists of experimental
data.
There are approximately 3.5 tunnelling periods in 40 modulation
periods in the theoretical curve.
Figure \ref{chap12fig9} shows the mean momentum as a function of
the interaction time with the standing wave for modulation
parameter $\varepsilon=0.30$, scaled well depth $\kappa=1.82$,
modulation frequency $\omega/2\pi=222$ kHz and phase shift
$\varphi=0.21\cdot2\pi$. For these parameters the tunnelling
frequency is larger than for the parameters shown in Fig.
\ref{chap12fig15}.

The simulation shows good agreement with the experiment. However,
the theoretical mean momentum tunnelling amplitude is larger than
the one measured in the experiment and the theoretical tunnelling
frequency for this set of parameters is slightly larger than the
experimentally measured one. Figure \ref{chap12fig10} illustrates that one can achieve much
better agreement between experiment and simulation if one of the
parameters is varied inside the experimental regime of
uncertainty. The theoretical curve in this figure is obtained
using the same parameters as in Fig. \ref{chap12fig9}, but the
modulation parameter $\varepsilon$ is reduced from 0.30 to 0.28.
The tunnelling amplitude and frequency is now very similar to the
experimental results. Note that the experimental data is not
centered at zero momentum. This is also the case in the
theoretical simulations for both $\varepsilon =0.28$ and
$\varepsilon=0.30$. The mean momentum curve appears much more
sinusoidal than the one for $\varepsilon=0.29$, $\kappa=1.66$,
$\omega /2\pi=250$ kHz and $\varphi=0.21\cdot2\pi$ (Fig.
\ref{chap12fig8}). This could imply that the initial wave function
has support on fewer Floquet states.

\subsection{Stroboscopic evolution of the system energies \label{stroben}}

\begin{figure}[ptbh] \centering
\includegraphics[width=8cm,keepaspectratio]{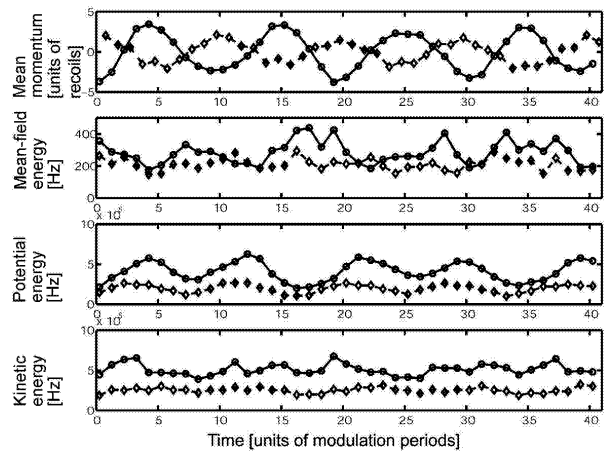}%
\caption{Stroboscopic evolution of mean-field energy, the kinetic
energy and the potential energy. The energies are given in Hz
(energy is scaled with Planck's constant). For comparison the mean
momentum is also shown as a function of the interaction time with
the standing wave (in units of modulation periods). The solid
(circles) and dashed (diamonds) curves correspond to the two
different interaction times $n+0.25$ and $n+0.75$ modulation
periods, respectively. The evolution is plotted for modulation
parameter $\varepsilon=0.29$, scaled well depth $\kappa=1.66$,
modulation frequency $\omega/2\pi=250$ kHz and a phase shift
$\varphi=0.21\cdot2\pi$.}%
\label{chap12fig25}%
\end{figure}

Calculating the expectation values of the relevant system energies
can give important information about the relevant energy scales
and it might also help to obtain a deeper insight into the
stroboscopic evolution of a Bose-Einstein condensate in a
periodically modulated potential. Figure \ref{chap12fig25} shows
the energy expectation values for the mean-field energy, the
potential energy and the kinetic energy given in Hz (scaled by
Planck's constant). For comparison the stroboscopic mean momentum expectation values
are also shown. The energy and momentum expectation values are
plotted stroboscopically and the solid (circles) and dashed
(diamonds) curves correspond to the two different interaction
times $n+0.25$ and $n+0.75$ modulation periods, respectively.
Figure \ref{chap12fig25} is plotted for modulation parameter
$\varepsilon=0.29$, scaled well depth $\kappa=1.66$, modulation
frequency $\omega/2\pi=250$ kHz and a phase shift of $\varphi
=0.21\cdot2\pi$ which corresponds to Fig. \ref{chap12fig8}. The
mean-field energy is three orders of magnitude smaller than the
potential or the kinetic energy. While the kinetic energy does not
show a distinct oscillation, an oscillation is clearly visible in
the stroboscopic potential energy evolution. This oscillation
frequency is not equal to the tunnelling frequency but it is
smaller as can be seen in Fig. \ref{chap12fig25}. One obtains a
period of approximately 8.9 modulation periods compared to a
tunnelling period of approximately 10.0 modulation periods.
Considering the energy scales one should note that this
oscillation is also clearly visible in the stroboscopic evolution
of the total energy of the atoms. The origin of this oscillation
is not known yet and will be the subject of future investigation.

\subsection{Floquet analysis for some experimental
parameters\label{Floquetanal}}

Time and spatial periodicity of the Hamiltonian allow utilization
of the Bloch and Floquet theorems \cite{Mouchet2001,Ashcroft76}.
Because of the time periodicity, there still exists eigenstates of
the evolution operator over one period (Floquet theorem).  Its
eigenvalues can be written in the
form~$e^{-i2\pi\phi_n\tau/\kbar}$ where~$\phi_n$ is called the
quasienergy of the states. $n$ is a discrete quantum number. Due
to the spatial~$\lambda/2$-periodicity, in addition to~$n$, these
states are labeled by a continuous quantum number, the so called
quasi-momentum~$\vartheta\in\left[
-2\pi/\lambda,2\pi/\lambda\right[$ (Bloch theorem). The
quasienergy spectrum~$\phi_n(\vartheta)$ is therefore made of
bands labelled by~$n$ (see for instance Fig. \ref{chap12fig4}).
More precisely, the states can be written as
\begin{equation}
\left|  \phi_{n,\vartheta}\left(  \tau\right)  \right\rangle
=e^{\left( -i2\pi\phi_{n}\left(  \vartheta\right)  \tau/ \kbar
\right)  }e^{\left(  -i\vartheta q\right) }\left|
\psi_{n,\vartheta}\left(  \tau\right)  \right\rangle
\label{Floq}%
\end{equation}
where $\left\{  \left|  \psi_{n,\vartheta}\left(  \tau\right)
\right\rangle \right\}  $ is now strictly periodic in space and
time (i.e. not up to a phase).

The evolution of the initial atomic wave function can be easily
computed from its expansion on the~$\left|
\phi_{n,\vartheta}\left( 0\right)  \right\rangle$ once the Floquet
operator has been diagonalized. The $\left|
\psi_{n,\vartheta}\left(  \tau\right)  \right\rangle$ are the
eigenstates of the modified Floquet-Bloch Hamiltonian
\begin{equation}
\mathcal{H}=\left(  p+\kbar \vartheta\right)
^{2}/2+2\kappa(1-2\varepsilon\sin\tau)\sin^{2}(q/2)
\end{equation}
subjected to strictly periodic space-time boundary conditions.

Dominant Floquet states may be determined by calculating the inner
product of the Floquet states with the initial atomic
wavefunction. To obtain a phase space representation of Floquet
states in momentum and position space one can calculate the Husimi
or Q-function. It is defined as
\begin{equation}
  Q\left(q,p,\tau\right)
  =\frac{1}{2\pi\kbar}
    \left|  \left\langle q+ip|\phi\right\rangle \right|  ^{2}%
\end{equation}
where $|q+ip\rangle$ is the coherent state of a simple harmonic
oscillator with frequency $\omega_{0}$ chosen as $\sqrt{\kappa}$
in scaled units. The position representation $\left\langle
q'|q+ip\right\rangle $ of the coherent state
\cite{Klauder85,Cohen77}\ is given by
\begin{equation}
\left\langle q'|q+ip\right\rangle
=\left(  \frac{\omega_{0}%
}{\pi\kbar}\right)^{1/4}\exp\left\{  -\left[  \frac{q'-q}{2\Delta
}\right]  ^{2}+ip\frac{q}{\kbar}\right\}
\end{equation}
up to an overall phase factor and where
$\Delta=\sqrt{\frac{\kbar}{2\omega_{0}}}$. Floquet analysis will
be shown for some of the experimental parameters which were
presented in the previous section.
\begin{figure}[ptbhptbh]
\centering
\includegraphics[width=6cm,keepaspectratio]{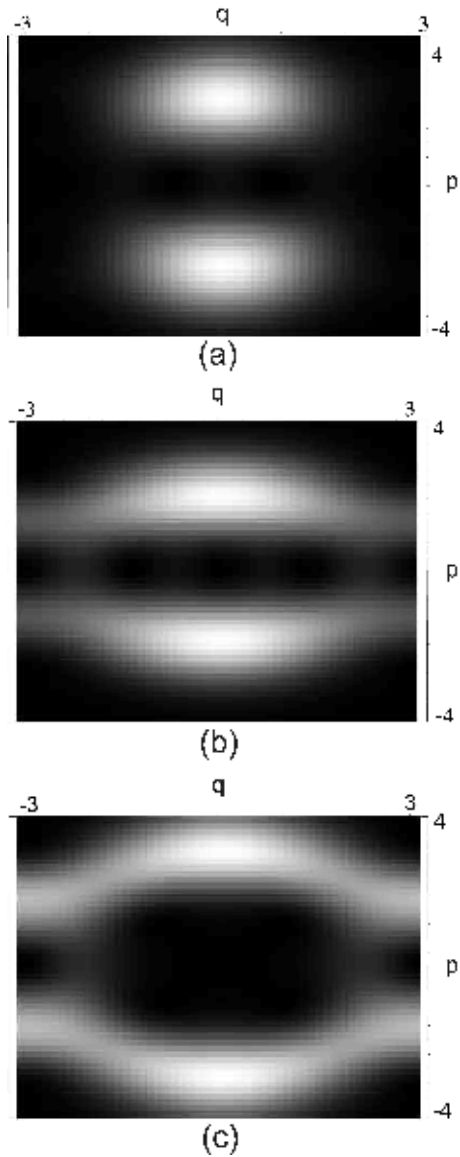}\caption{Phase
space representations of two Floquet states for~$\vartheta=0$ that
are involved in the dynamical tunnelling are shown in (a) and (b).
The Floquet states correspond to the experimental parameters:
modulation parameter $\varepsilon =0.29$, scaled well depth
$\kappa=1.66$ and modulation frequency $\omega /2\pi=250$ kHz,
which were utilized to obtain the experimental results shown in
Fig. \ref{chap12fig8}. A third state, shown in (c) also has
significant overlap with the initial
experimental state.}%
\label{chap12fig2}%
\end{figure}Figure \ref{chap12fig2} shows contour plots of the Husimi
functions of two Floquet states with opposite parity ((a), (b))
for these parameters whose presence allows dynamical tunnelling to
occur. Both of them are approximately localized on the classical
period-1 regions of regular motion and they were selected so that
the initial atomic wave function has significant overlap with them
(26\% and 44\%, respectively). The initial experimental state has
also significant overlap (22\%) with a third state that is shown
in Fig. \ref{chap12fig2}(c). The overlap is calculated using a
coherent state that is centered on the periodic region of regular
motion, which is a good approximation of the initial experimental
state. The quasi-eigenenergy spectrum for this set of parameters
is shown in Fig. \ref{chap12fig4}.
\begin{figure}[ptbh] \centering
\includegraphics[width=8cm,keepaspectratio]{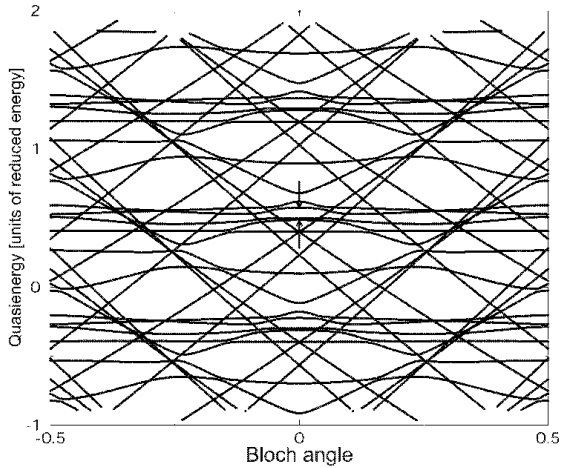}%
\caption{Quasi-eigenenergy spectrum for parameters modulation
parameter $\varepsilon=0.29$, scaled well depth $\kappa=1.66$ and
modulation frequency $\omega/2\pi=250$ kHz. The quasi-eigenenergy
is measured in reduced energy units and it is a function of the
Bloch angle. The Bloch angle is equal to the quasi-momentum
multiplied with the spatial period $\lambda/2$ of the lattice.
Each line corresponds to one Floquet state labelled by~$n$ (see
Eq.\ref{Floq}). The quasi-eigenenergy of most Floquet states is
strongly dependent on the quasi-momentum but it is not the case
for the states of the tunnelling doublet. Recall that the
quasi-momentum~$\vartheta$ is proportional to the phase taken by a
state when spatially translated by $\lambda/2$. In other words,
fixing $\vartheta$ imposes on the states some conditions on the
boundary of one elementary cell $[q,q+\lambda/2]$. If a state is
localized deep inside a cell, changing the boundary conditions
(i.e. $\vartheta$) will not affect the state so much and the
corresponding quasienergies appear as curves that are
approximately parallel to the $\vartheta$ axis. However, one
should expect a strong $\vartheta$-dependence for a state that
spread over at least~$\Delta q\sim\lambda/2$.}%
\label{chap12fig4}%
\end{figure}The quasi-eigenenergies are plotted as a function of the
quasi-momentum (Bloch angle). Each line corresponds to one Floquet
state labelled by~$n$ (see Eq.\ref{Floq}). The arrows point to the
two Floquet states shown in Fig. \ref{chap12fig2}(a) and (b) that
correspond to the tunnelling splitting in Eq.
~\ref{eq:splitting}. Most of the other lines correspond to Floquet
states which lie in the classical chaotic phase space region.
Examples of phase space representations of such Floquet states can
be found in reference \cite{Mouchet2001}. Note that the spectrum
is $2\pi\kbar/T$ periodic in quasienergies. For quasi-momentum
$\vartheta=0$ the splitting between the two states, that are shown
in Fig. \ref{chap12fig2} is approximately 0.08 in reduced units
(energy in frequency units, [energy$/\left( 2\pi\kbar\right) $]).
With a modulation frequency $\omega/2\pi=250$ kHz one obtains a
scaled Planck's constant $\kbar=0.8$, therefore the tunnelling
period $T_{tun}$ which follows from Eq.  \ref{period} is~10 which
is in good agreement with the experiment.

The quasi-momentum plays a significant role in the experiments.
The quasi-momentum $\vartheta$ is approximately equal to the
relative velocity $v$ between the wave packet (before the lattice
is turned on) and the lattice, $v=\vartheta/m$
\cite{Denschlag2002} if the standing wave is adiabatically turned
on.  It was found in~\cite{Mouchet2001} that it is of importance
to populate a state whose quasi-momentum average is equal to zero.
Moreover, quasi-momentum spread is also of importance. It has been
shown in~\cite{Mouchet03a} that if the thermal velocity
distribution is too broad, then the tunnelling oscillation
disappears. As can be seen in Fig.  \ref{chap12fig4} the
quasi-eigenenergies of the two contributing Floquet states depend
on the quasi-momentum (or Bloch angle). The tunnelling period
(which is inversely proportional to the separation between these
two quasi-eigenenergies) depends on the quasi-momentum.  Using a
thermal atomic cloud one obtains a statistical ensemble of many
quasi-momenta as they initially move in random directions with
respect to the optical lattice.  Atoms localized in individual
wells can be described by a wave packet in the plane wave basis
and therefore they are characterized by a superposition of many
quasi-momenta. The resulting quasi-momentum width washes out the
tunnelling oscillations (see \cite[Fig.~3]{Mouchet03a}).  In fact
in another experiment Steck et al. \cite{Steck2001} found that the
amplitude of the mean momentum oscillations resulting from a
tunnelling process between two librational islands of stability
decreased when the initial momentum width of the atomic cloud was
increased.
\begin{figure}[ptbh] \centering
\includegraphics[width=6cm,keepaspectratio]{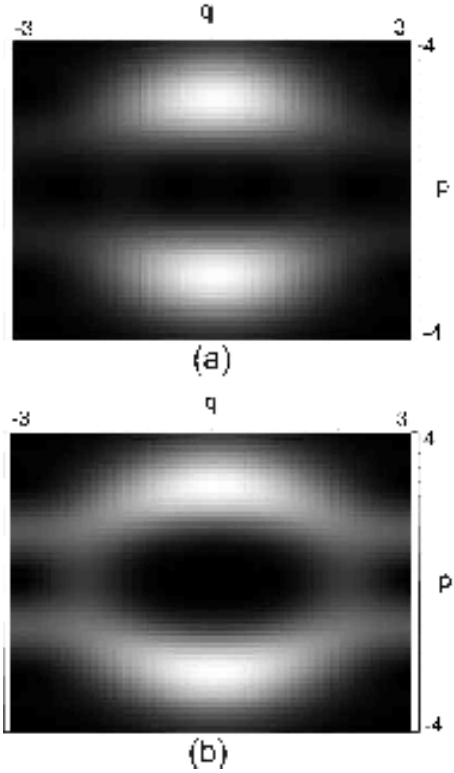}\caption{Phase
space representations of two Floquet states whose presence can lead to
the occurrence of dynamical tunnelling. The Floquet states correspond
to the experimental parameters: modulation parameter
$\varepsilon=0.30$, scaled well depth $\kappa=1.82$ and modulation
frequency $\omega/2\pi=222$ kHz, which were utilized to obtain the
experimental results shown in Fig. \ref{chap12fig9}.}%
\label{chap12fig3}%
\end{figure}

Figure \ref{chap12fig3} shows contour plots of Husimi functions
for the two dominant Floquet states for the modulation parameter
$\varepsilon=0.30$, scaled well depth $\kappa=1.82$ and modulation
frequency $\omega/2\pi=222$ kHz, which corresponds to experimental
results shown in Fig. \ref{chap12fig9}. The states are selected to
have maximum overlap with the initial wave packet (38\% and 44\%).
In contrast to the experimental results shown in Fig.
\ref{chap12fig8} there are only two dominant Floquet states. The
quasi-eigenenergy spectrum for this set of parameters (not shown)
reveals a level splitting of approximately 0.15 in reduced units,
the calculated tunnelling period is 6 modulation periods which is
in good agreement with the experiment.

When comparing the stroboscopic evolution of the mean momentum
shown in Fig. \ref{chap12fig8} \ (modulation parameter
$\varepsilon=0.29$, scaled well depth $\kappa=1.66$ and modulation
frequency $\omega/2\pi=250$ kHz) with the one shown in Fig.
\ref{chap12fig9} (modulation parameter $\varepsilon=0.30$, scaled
well depth $\kappa=1.82$ and modulation frequency
$\omega/2\pi=222$ kHz), one finds that it is less sinusoidal. This
can be explained in the Floquet picture. While there are three
dominant Floquet states for the first case (Figs.
\ref{chap12fig8}, \ref{chap12fig2}) (three Floquet states with
significant overlap with the initial experimental state), there
are only two dominant Floquet states for the second case (Figs.
\ref{chap12fig9}, \ref{chap12fig3}) resulting in a more sinusoidal
tunnelling oscillation.

\subsection{Loading analysis of the Floquet superposition state}

\begin{figure}[ptbh]
\begin{center}
\includegraphics[width=8cm,keepaspectratio]{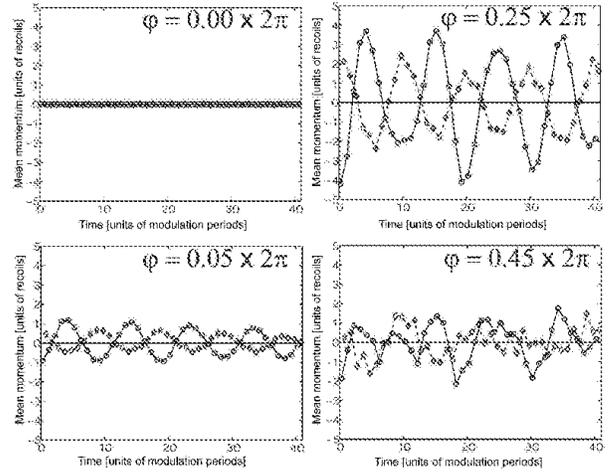}
\end{center}
\caption{Mean atomic momentum as a function of the interaction time with the
standing wave plotted for a range of the initial phase shift $\varphi$ of the
standing wave. The solid line (circles) corresponds to an interaction time
with the standing wave of $n+0.25$ modulation periods and the dashed line
(diamonds) corresponds to an interaction time of $n+0.75$ modulation periods.}%
\label{chap12fig26}%
\end{figure}

The initial atomic wave packet is localized around the classical
period-1 region of regular motion by inducing a sudden phase shift
to the standing wave. This enables the observation of dynamical
tunnelling. In the Floquet picture the observation of dynamical
tunnelling requires that the initial state has support on only a
few dominant Floquet states, preferably populating only two with a
phase space structure as shown in Fig.~\ref{chap12fig2}.
 Optimizing the overlap of the
initial state with these Floquet ``tunnelling'' states should maximize the
observed tunnelling amplitude. Here we carry out analysis confirming this
prediction. This corresponds to optimizing the overlap of the initial
experimental state with the period-1 regions of regular motion. Figure
\ref{chap12fig26} shows the mean atomic momentum as a function of the
interaction time with the standing wave which is plotted for a range of the
initial phase shift $\varphi$ of the standing wave. The solid line (circles) corresponds to an interaction time with
the standing wave of $n+0.25$ modulation periods and the dashed
line (diamonds) corresponds to an interaction time of $n+0.75$
modulation periods. The simulations are made using the
Gross-Pitaevskii equation for modulation parameter
$\varepsilon=0.29$, scaled well depth $\kappa=1.66$, modulation
frequency $\omega/2\pi=250$ kHz which corresponds to Fig.
\ref{chap12fig8}. A phase shift $\varphi=0$ corresponds to
localizing the wave packet exactly at the bottom of the well and
$\varphi=\pi$ corresponds to localizing it exactly at the maximum
of the standing wave well. Symmetry dictates that no tunnelling
oscillation can occur for these two loading phases. This is also
shown in Fig. \ref{chap12fig26} (the mean momentum curve for
$\varphi=\pi$ is not shown but it is the same as for $\varphi=0$).
The amplitude of the tunnelling oscillations changes strongly when
the initial phase shift $\varphi$ of the standing wave is changed.
The best overlap with the tunnelling Floquet states is obtained
for the phase shift $\varphi$ somewhere in between $0.25\times
2\pi$ and $0.30\times2\pi$ which corresponds to placing the wave
packet half way up the standing wave well. This result is in good
agreement with the structure of the tunnelling Floquet states as
shown in Fig.~\ref{chap12fig2}.  When changing $\varphi$ there is
no change in the observed tunnelling period. This is to be
expected as mainly the loading efficiency for the ``tunnelling''
Floquet states varies when $\varphi$ is varied. It should be noted
that the simulations are carried out for a relatively large scaled
Planck's constant ($\kbar=0.8$) which means that the wave packet
size is rather large compared to characteristic classical phase
space features like the period-1 regions of regular motion. This
analysis shows that the tunnelling amplitude sensitively depends
on the loading efficiency of the tunnelling Floquet states and
that there is a smooth dependency of the tunnelling amplitude on
this loading parameter.

\section{Parameter dependency of the tunnelling
oscillations}\label{three}

The scaled parameter space for the dynamics of the system is given
by the scaled well depth $\kappa$ and the modulation parameter
$\varepsilon.$ Both parameters will significantly change the
structure and number of the contributing Floquet states. It has
been found that a strong sensitivity of the tunnelling frequency
on the system parameters is a signature of chaos-assisted
tunnelling \cite{Mouchet2001,Mouchet03a} where a third state
associated with the classical chaotic region interacts with the
tunnelling Floquet states. A Floquet state that is localized
inside a region of regular motion that surrounds another resonance
can also interact with the tunnelling doublet (two tunnelling
Floquet states), this phenomenon is known as resonance-assisted
tunnelling \cite{Brodier2001}.

Comprehensively exploring the parameter space and its associated
phenomena is out of the scope of this article. Instead an analysis
associated with our experiments will be presented here showing one
scan of the scaled well depth $\kappa$ and another scan of of the
modulation parameter $\varepsilon$ around the experimental
parameter regime. The results are shown in the form of plots of
the mean momentum as a function of the interaction time with the
modulated standing wave. The solid line (circles) corresponds to
an interaction time with the standing wave of $n+0.25$ modulation
periods and the dashed line (diamonds) corresponds to an
interaction time of $n+0.75$ modulation periods.

\begin{figure}[ptbh] \centering
\includegraphics[width=7cm,keepaspectratio]{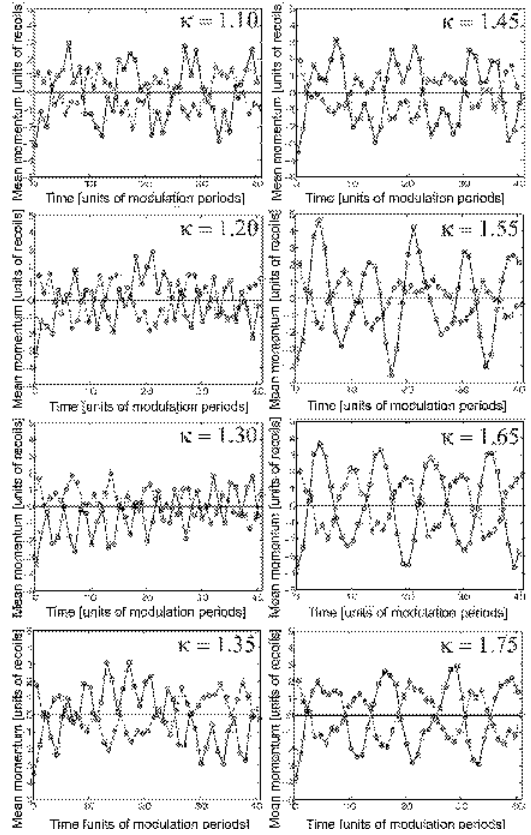}\caption{Mean
momentum as a function of the interaction time with the modulated
standing wave for different values of the scaled well depth $\kappa$
(modulation parameter $\varepsilon=0.29$, modulation frequency
$\omega/2\pi=250$ kHz and phase shift $\varphi=0.21\cdot2\pi$). The
solid line (circles) corresponds to an interaction time with the
standing wave of $n+0.25$ modulation periods and the dashed line
(diamonds) corresponds to an interaction time of $n+0.75$ modulation
periods.}%
\label{chap12fig27}%
\end{figure}

Figure \ref{chap12fig27} shows the scaled well depth $\kappa$
being varied from 1.10 to 1.75.
\begin{figure}[ptbh] \centering
\includegraphics[width=7cm,keepaspectratio]{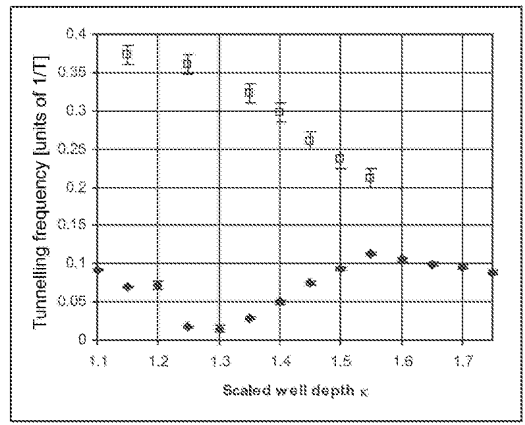}%
\caption{Tunnelling frequency as a function of the scaled well
depth $\kappa$ (modulation parameter $\varepsilon=0.29$,
modulation frequency $\omega /2\pi=250$ kHz and phase shift
$\varphi=0.21\cdot2\pi$) taken from Fig. \ref{chap12fig27}. Diamonds and squares denote low and high frequency components, respectively.}%
\label{kappaperiod}%
\end{figure}
\begin{figure*}[ptbh]
\centering
\includegraphics[width=13cm,keepaspectratio]{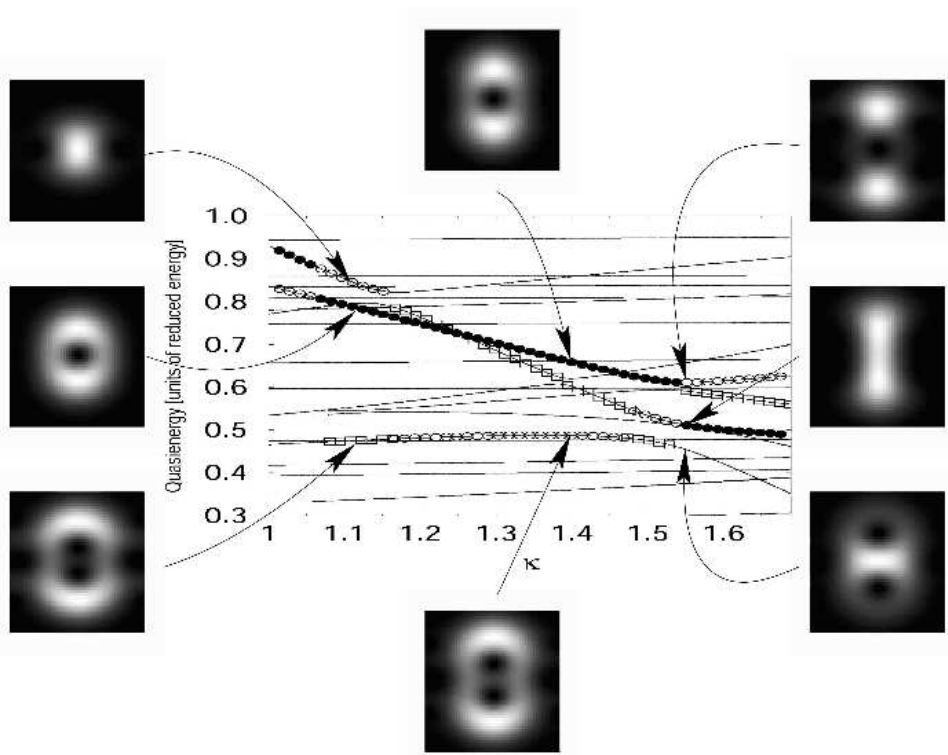}
\caption{Floquet spectrum as a function of the scaled well depth
$\kappa$ (modulation parameter $\varepsilon=0.29$, modulation
frequency $\omega/2\pi=250$ kHz and phase shift
$\varphi=0.21\cdot2\pi$. The Floquet states with maximum overlap
are marked black bullets, white bullets are used for second most
overlap and white squared bullets show third most overlap. Phase
space representations (Husimi functions) of some of the these
Floquet states for different values of $\kappa$ are
also shown.}%
\label{kappaFloquetspectrum}%
\end{figure*}
The other parameters are held constant (modulation parameter
$\varepsilon=0.29$, modulation frequency $\omega/2\pi=250$ kHz and
phase shift $\varphi=0.21\cdot2\pi$). The momentum distribution
evolution is shown to illustrate intricate changes in the
tunnelling dynamics when the parameters are varied. Both the
tunnelling frequency spectrum and the tunnelling amplitude are
strongly dependent on $\kappa.$ Often one can see more than just
one dominant tunnelling frequency. For example, for $\kappa=1.35$
the two dominant tunnelling frequencies which contribute to the
tunnelling oscillation have a period of approximately 3.9
modulation periods and approximately 34 modulation periods. In the
interval of approximately $\kappa=1.5$ and $\kappa=1.8$ the
tunnelling oscillations have a more sinusoidal shape indicating
the presence of only approximately two dominant tunnelling states.
In this interval the tunnelling frequency is peaked at
$\kappa=1.55$ with a tunnelling period of approximately 8.9
modulation periods. In order to analyze further the behavior of
the tunnelling frequency, we show the low frequency (diamonds) and
the high frequency component (squares) of the tunnelling frequency
as a function of the scaled well depth $\kappa$ in
Fig.~\ref{kappaperiod}. The high frequency component is plotted
only for values of $\kappa$ where it is visible in the
stroboscopic momentum evolution shown in Fig. \ref{chap12fig27}.
The occurrence of multiple tunnelling frequencies results from the
presence of more than two dominant Floquet states. The minimum of
the low frequency component of the tunnelling frequency is due the
the level crossing of two contributing Floquet states (see squares
and filled circles at $\kappa=1.25$ in Fig.
\ref{kappaFloquetspectrum}). The error bars result from the
readout uncertainty of the tunnelling period from the simulations.
This analysis reveals some of the intricate features of the
tunnelling dynamics. Instead of a smooth parameter dependency a
distinct rise and fall (in the low frequency component) in the
tunnelling frequency versus scaled well depth appears. The
tunnelling frequency minimum is centered at $\kappa \approx1.3$.
Note that a rise and fall in the tunnelling frequency is often
understood as a signature of chaos-assisted tunnelling. However,
it is not a sufficient criterion for chaos-assisted tunnelling.
One needs to choose an approximately ten times smaller scaled
Planck's constant to use the terminology of chaos-assisted
tunnelling \cite{Mouchet03a}. The size of the Floquet states is
given by the scaled Planck's constant. If the states are much
larger than phase space features like regions of regular motion,
then it is impossible to make a classification of Floquet states
as chaotic or regular, which is needed for chaos-assisted
tunnelling.

Another interesting feature of dynamical tunnelling can be derived
from Fig.~\ref{kappaperiod}. It shows that for a certain parameter
regime the tunnelling frequency decreases with decreasing scaled
well depth. This is the contrary of what one would expect for
spatial energy barrier tunnelling. This feature has been also
observed in recent experiments \cite{Steck2002}.
 Floquet spectra provide alternative means of
analyzing the tunnelling dynamics. The dependence of the
tunnelling frequency on the scaled well depth $\kappa$ can be
understood using the appropriate Floquet spectrum. Figure
\ref{kappaFloquetspectrum} shows the quasi-eigenenergies of
different Floquet states as a function of the scaled well depth
$\kappa$. The Floquet states with maximum overlap are marked with
bullets. Figure \ref{kappaFloquetspectrum} also shows phase space
representations (Husimi functions) of some of these Floquet states
for different values of $\kappa$.  The shape and structure of
these Floquet states depend on the value of the scaled well depth.
In fact Floquet states can undergo bifurcations. This may be seen
as the quantum analogue of classical phase bifurcation. Classical
phase space bifurcations have been reported in ref.
\cite{Hensinger2001a}. In this case the shapes of the Floquet
states change in such a way that different Floquet states have
non-negligible overlap with the initial experimental state as the
scaled well depth $\kappa$ is varied. The separation between the
quasi-eigenenergies of the two states with maximum overlap will
determine a dominant tunnelling frequency. Note that often more
than two states have relevant overlap with the initial
experimental states which leads to the occurrence of multiple
tunnelling frequencies.

\begin{figure}[ptbh]
\centering
\includegraphics[width=8cm,keepaspectratio]{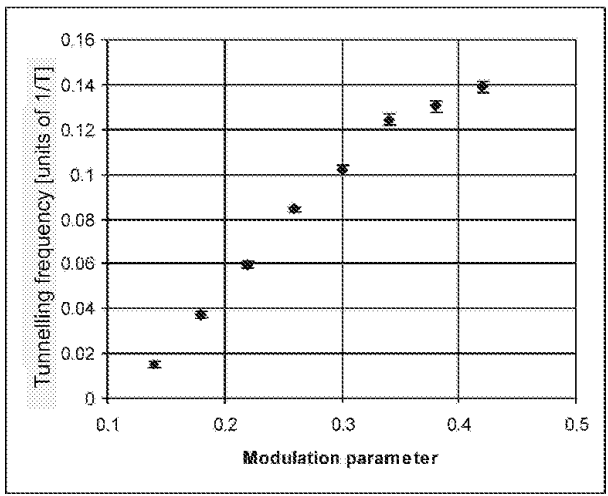}%
\caption{Tunnelling frequency as a function of the modulation
parameter $\varepsilon$ (scaled well depth $\kappa=1.66,$
modulation frequency $\omega/2\pi=250$ kHz and phase shift
$\varphi=0.21\cdot2\pi$). The error bars result from the readout
uncertainty of the tunnelling oscillations from the simulations.
The high frequency component that
is present in the data for $0.00<\varepsilon<0.18$ is not shown.}%
\label{chap12fig28b}%
\end{figure}

Figure \ref{chap12fig28b} shows effects of a variation of the
modulation parameter $\varepsilon.$
To obtain these simulations all other parameters are held constant
(scaled well depth $\kappa=1.66,$ modulation frequency
$\omega/2\pi=250$ kHz and phase shift $\varphi=0.21\cdot2\pi$).
For smaller values of $\varepsilon$ the corresponding classical
phase space is mainly regular. There is no distinct oscillation at
$\varepsilon=0.10$ which is centered at zero momentum. Distinct
tunnelling oscillations start to occur at $\varepsilon=0.14$ with
a gradually increasing tunnelling frequency. While there is a
tunnelling period of approximately 67 modulation periods at
$\varepsilon=0.14,$ the tunnelling period is only approximately
7.7 modulation periods at $\varepsilon=0.38.$ For larger values of
the modulation parameter $\varepsilon$ the oscillations become
less sinusoidal indicating the presence of an increasing number of
dominant Floquet states. The error bars result from the readout
uncertainty of the tunnelling period from the simulations.

\section{Moving the quantum system towards the classical
limit}\label{four}

A fundamental strength of the experiments which are discussed here
is that they are capable of exploring the transition of the
quantum system towards the classical limit by decreasing the
scaled Planck's constant $\kbar$.

\begin{figure}[ptbh] \centering
\includegraphics[width=7cm,keepaspectratio]{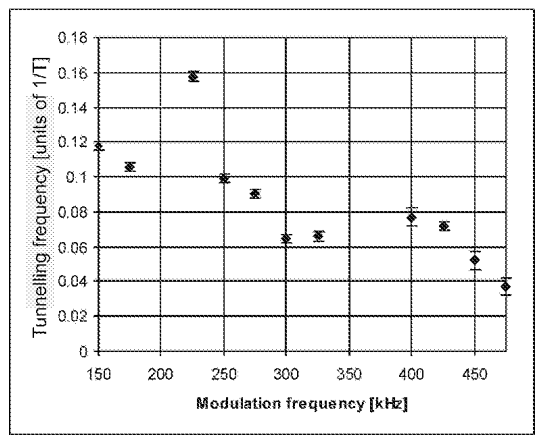}%
\caption{Dominant tunnelling frequency as a function of the
modulation frequency $\omega/2\pi$ for modulation parameter
$\varepsilon=0.29,$ scaled well depth $\kappa=1.66$ and phase
shift $\varphi=0.21\cdot2\pi$ which corresponds to experimental
results shown in Fig. \ref{chap12fig8}. As expected, the overall
tendency consists of a decreasing of the tunnelling frequency as
the modulation frequency increases and therefore as the scaled
Planck's constant decreases, however the transition is not as
smooth as one might expect.}
\label{omegaperiod}%
\end{figure}

Here a quantum system with mixed phase space exhibiting
classically chaotic and regular regions of motion is moved towards
the classical limit. By adjusting the scaled Planck's constant
$\kbar$ of the system, the wave and particle character of the
atoms can be probed although some experimental and numerical
restrictions limit the extent of this quantum-classical probe.
This should enhance our understanding of nonlinear dynamical
systems and provide insight into their quantum and classical
origin. It is out of the scope of this paper to present anything
more than a short analysis relevant to the experimental results.
The quantum-classical borderland is analyzed by considering the
mean momentum as a function of the interaction time with the
modulated standing wave for different values of the scaled
Planck's constant $\kbar$.  Figure \ref{omegaperiod} shows results
for modulation parameter $\varepsilon =0.29,$ scaled well depth
$\kappa=1.66$ and phase shift $\varphi=0.21\cdot 2\pi$ which
corresponds to experimental results shown in Fig.
\ref{chap12fig8}. The scaled Planck's constant $\kbar$ is varied
by adjusting the modulation frequency $\omega/2\pi$ and leaving
the scaled well depth $\kappa,$ the modulation parameter
$\varepsilon$ and the initial phase shift $\varphi$ constant.
Results are shown for the scaled Planck's constant $\kbar$ ranging
from 0.40 ($\omega/2\pi=500$ kHz) to 1.33 ($\omega/2\pi=150$ kHz).
The error bars result from the readout uncertainty of the
tunnelling period from the simulations.
As shown in previous work \cite{Hensinger2001b}, the momentum of
the regions of regular motion is proportional to the modulation
frequency. Note that this is a purely classical feature that would
disappear if the scaled momentum would be plotted instead of the
real momentum. Intuitively one would expect the tunnelling
frequency to decrease as the system becomes more classical.
However, a different quite surprising phenomenon occurs.
Considering the modulation frequency interval between 225 kHz and
325 kHz the tunnelling frequency gradually decreases as the system
becomes more classical. The same applies for the interval between
150 kHz to 175 kHz and the interval from 425 kHz to 475 kHz.
Qualitative changes occur at approximately 200 kHz, 350 kHz (375
kHz) and 500 kHz where the tunnelling oscillation shows first
multi-frequency contributions and is then significantly increased
as the effective Planck's constant is further decreased.  A more
detailed analysis will be subject of future study. The three
intervals where the tunnelling frequency decreases appear as three
arms in the graph. At $\omega/2\pi=200$ kHz, 350 kHz and 375 kHz
the system cannot be well described using one dominant tunnelling
frequency, therefore no data is shown at these frequencies.

The results of this first analysis indicate that the transition
from quantum to classical physics contains many fascinating
details to be explored. The results for the driven pendulum in
atom optics shows that the transition from quantum to classical
dynamics is not smooth.

\section{Conclusion}\label{five}

Recent dynamical tunnelling experiments \cite{Hensinger2001c} were
analyzed using simulations of the Gross-Pitaevskii equation and the
corresponding Floquet theory. The main features of the experiments can
be explained by a two or three state framework that is provided by
Floquet theory. We have identified the relevant Floquet states and
shown their Husimi functions. Note that the mean-field interaction was
negligible in the experiments.  Tunnelling period and amplitude are in
good agreement with GP simulation and Floquet theory.

An analysis of the parameter space has shown that tunnelling frequency
is strongly dependent on the system parameters scaled well depth
$\kappa,$ modulation parameter $\varepsilon$ and the scaled Planck's
constant $\kbar$. While we find an approximately linear dependence of
the tunnelling frequency on the modulation parameter $\varepsilon$ for
the set of experimental parameters, there is a distinct spike in the
tunnelling frequency as a function of the scaled well depth $\kappa.$

However this cannot be interpreted as a signature of
chaos-assisted tunnelling essentially because the states cannot be
clearly identified with chaotic or regular regions of the
classical phase space, therefore the notion of chaos-assisted
tunnelling is difficult to use in this context. We note that it is
important to decrease the scaled Planck's constant by at least an
order of magnitude \cite{Mouchet03a} for an observation of
chaos-assisted tunnelling to be non-ambiguous.  We have simulated
when the system is moved towards the classical limit. A
bifurcation-like behavior results which can be understood in terms
of quantum bifurcations of the contributing Floquet states.

\section{Acknowledgements}

This work was supported by the Australian Research Council. The
experimental results that are shown here were accomplished by W.
K. Hensinger, H. H\"{a}ffner, A. Browaeys, N. R. Heckenberg, K.
Helmerson, C. McKenzie, G. J. Milburn, W. D. Phillips, S. L.
Rolston, H. Rubinsztein-Dunlop, and B. Upcroft in the context of
reference \cite{Hensinger2001c} and we would like to thank the
authors. The corresponding author W. K. H. would like to
acknowledge interesting discussions with William D. Phillips and
Steven Rolston during the preparation of theoretical results
presented in this article. W. K. H. would like to thank NIST and
Laboratoire Kastler-Brossel for hospitality during part of the
work for this paper. Laboratoire Kastler Brossel de
l'Universit\'{e} Pierre et Marie Curie et de l'Ecole Normale
Sup\'{e}rieure is Unit\'e Mixte de Recherche 8552 du CNRS.
Laboratoire de math\'{e}matiques et de physique th\'{e}orique de
l'Universit\'{e} Fran\c{c}ois Rabelais is Unit\'e Mixte de
Recherche 6083 du CNRS.


\begin{thebibliography}{22}
\expandafter\ifx\csname
natexlab\endcsname\relax\def\natexlab#1{#1}\fi
\expandafter\ifx\csname bibnamefont\endcsname\relax
  \def\bibnamefont#1{#1}\fi
\expandafter\ifx\csname bibfnamefont\endcsname\relax
  \def\bibfnamefont#1{#1}\fi
\expandafter\ifx\csname citenamefont\endcsname\relax
  \def\citenamefont#1{#1}\fi
\expandafter\ifx\csname url\endcsname\relax
  \def\url#1{\texttt{#1}}\fi
\expandafter\ifx\csname
urlprefix\endcsname\relax\def\urlprefix{URL }\fi
\providecommand{\bibinfo}[2]{#2}
\providecommand{\eprint}[2][]{\url{#2}}

\bibitem[{\citenamefont{Hensinger
  et~al.}(2001{\natexlab{a}})\citenamefont{Hensinger, H{\"a}ffner, Browaeys,
  Heckenberg, Helmerson, McKenzie, Milburn, Phillips, Rolston,
  Rubinsztein-Dunlop et~al.}}]{Hensinger2001c}
\bibinfo{author}{\bibfnamefont{W.~K.} \bibnamefont{Hensinger}},
  \bibinfo{author}{\bibfnamefont{H.}~\bibnamefont{H{\"a}ffner}},
  \bibinfo{author}{\bibfnamefont{A.}~\bibnamefont{Browaeys}},
  \bibinfo{author}{\bibfnamefont{N.~R.} \bibnamefont{Heckenberg}},
  \bibinfo{author}{\bibfnamefont{K.}~\bibnamefont{Helmerson}},
  \bibinfo{author}{\bibfnamefont{C.}~\bibnamefont{McKenzie}},
  \bibinfo{author}{\bibfnamefont{G.~J.} \bibnamefont{Milburn}},
  \bibinfo{author}{\bibfnamefont{W.~D.} \bibnamefont{Phillips}},
  \bibinfo{author}{\bibfnamefont{S.~L.} \bibnamefont{Rolston}},
  \bibinfo{author}{\bibfnamefont{H.}~\bibnamefont{Rubinsztein-Dunlop}},
  \bibnamefont{et~al.}, \bibinfo{journal}{Nature}
  \textbf{\bibinfo{volume}{412}}, \bibinfo{pages}{52}
  (\bibinfo{year}{2001}{\natexlab{a}}).

\bibitem[{\citenamefont{Hensinger et~al.}(2003)\citenamefont{Hensinger,
  Heckenberg, Milburn, and Rubinsztein-Dunlop}}]{Hensinger03}
\bibinfo{author}{\bibfnamefont{W.~K.} \bibnamefont{Hensinger}},
  \bibinfo{author}{\bibfnamefont{N.~R.} \bibnamefont{Heckenberg}},
  \bibinfo{author}{\bibfnamefont{G.~J.} \bibnamefont{Milburn}},
  \bibnamefont{and}
  \bibinfo{author}{\bibfnamefont{H.}~\bibnamefont{Rubinsztein-Dunlop}},
  \bibinfo{journal}{J. Opt. B: Quantum Semiclass. Opt.}
  \textbf{\bibinfo{volume}{5}}, \bibinfo{pages}{R83} (\bibinfo{year}{2003}).

\bibitem[{\citenamefont{Steck et~al.}(2001)\citenamefont{Steck, Oskay, and
  Raizen}}]{Steck2001}
\bibinfo{author}{\bibfnamefont{D.~A.} \bibnamefont{Steck}},
  \bibinfo{author}{\bibfnamefont{W.~H.} \bibnamefont{Oskay}}, \bibnamefont{and}
  \bibinfo{author}{\bibfnamefont{M.~G.} \bibnamefont{Raizen}},
  \bibinfo{journal}{Science} \textbf{\bibinfo{volume}{293}},
  \bibinfo{pages}{274} (\bibinfo{year}{2001}).

\bibitem[{\citenamefont{Steck et~al.}(2002)\citenamefont{Steck, Oskay, and
  Raizen}}]{Steck2002}
\bibinfo{author}{\bibfnamefont{D.~A.} \bibnamefont{Steck}},
  \bibinfo{author}{\bibfnamefont{W.~H.} \bibnamefont{Oskay}}, \bibnamefont{and}
  \bibinfo{author}{\bibfnamefont{M.~G.} \bibnamefont{Raizen}},
  \bibinfo{journal}{Phys. Rev. Lett.} \textbf{\bibinfo{volume}{88}},
  \bibinfo{pages}{120406} (\bibinfo{year}{2002}).

\bibitem[{\citenamefont{Luter and Reichl}(2002)}]{Luter02}
\bibinfo{author}{\bibfnamefont{R.}~\bibnamefont{Luter}} \bibnamefont{and}
  \bibinfo{author}{\bibfnamefont{L.~E.} \bibnamefont{Reichl}},
  \bibinfo{journal}{Phys. Rev. A} \textbf{\bibinfo{volume}{66}},
  \bibinfo{pages}{053615} (\bibinfo{year}{2002}).

\bibitem[{\citenamefont{Averbukh et~al.}(2002)\citenamefont{Averbukh, Osovski,
  and Moiseyev}}]{Averbukh02}
\bibinfo{author}{\bibfnamefont{V.}~\bibnamefont{Averbukh}},
  \bibinfo{author}{\bibfnamefont{S.}~\bibnamefont{Osovski}}, \bibnamefont{and}
  \bibinfo{author}{\bibfnamefont{N.}~\bibnamefont{Moiseyev}},
  \bibinfo{journal}{Phys. Rev. Lett.} \textbf{\bibinfo{volume}{89}},
  \bibinfo{pages}{253201} (\bibinfo{year}{2002}).

\bibitem[{\citenamefont{Moore et~al.}(1995)\citenamefont{Moore, Robinson,
  Bharucha, Sundaram, and Raizen}}]{Moore95}
\bibinfo{author}{\bibfnamefont{F.~L.} \bibnamefont{Moore}},
  \bibinfo{author}{\bibfnamefont{J.~C.} \bibnamefont{Robinson}},
  \bibinfo{author}{\bibfnamefont{C.~F.} \bibnamefont{Bharucha}},
  \bibinfo{author}{\bibfnamefont{B.}~\bibnamefont{Sundaram}}, \bibnamefont{and}
  \bibinfo{author}{\bibfnamefont{M.~G.} \bibnamefont{Raizen}},
  \bibinfo{journal}{Phys. Rev. Lett.} \textbf{\bibinfo{volume}{75}},
  \bibinfo{pages}{4598} (\bibinfo{year}{1995}).

\bibitem[{\citenamefont{Dalfovo et~al.}(1999)\citenamefont{Dalfovo, Giorgini,
  Pitaevskii, and Stringari}}]{Dalfovo99}
\bibinfo{author}{\bibfnamefont{F.}~\bibnamefont{Dalfovo}},
  \bibinfo{author}{\bibfnamefont{S.}~\bibnamefont{Giorgini}},
  \bibinfo{author}{\bibfnamefont{L.~P.} \bibnamefont{Pitaevskii}},
  \bibnamefont{and}
  \bibinfo{author}{\bibfnamefont{S.}~\bibnamefont{Stringari}},
  \bibinfo{journal}{Rev. Mod. Phys.} \textbf{\bibinfo{volume}{71}},
  \bibinfo{pages}{463} (\bibinfo{year}{1999}).

\bibitem[{\citenamefont{Parkins and Walls}(1997)}]{Parkins97}
\bibinfo{author}{\bibfnamefont{A.~S.} \bibnamefont{Parkins}} \bibnamefont{and}
  \bibinfo{author}{\bibfnamefont{D.~F.} \bibnamefont{Walls}},
  \bibinfo{journal}{Phys. Rep.} \textbf{\bibinfo{volume}{303}},
  \bibinfo{pages}{1} (\bibinfo{year}{1997}).

\bibitem[{\citenamefont{Dyrting et~al.}(1993)\citenamefont{Dyrting, Milburn,
  and Holmes}}]{Dyrting93}
\bibinfo{author}{\bibfnamefont{S.}~\bibnamefont{Dyrting}},
  \bibinfo{author}{\bibfnamefont{G.~J.} \bibnamefont{Milburn}},
  \bibnamefont{and} \bibinfo{author}{\bibfnamefont{C.~A.}
  \bibnamefont{Holmes}}, \bibinfo{journal}{Phys. Rev. E}
  \textbf{\bibinfo{volume}{48}}, \bibinfo{pages}{969} (\bibinfo{year}{1993}).

\bibitem[{\citenamefont{Mouchet et~al.}(2001)\citenamefont{Mouchet, Miniatura,
  Kaiser, Gr\'{e}maud, and Delande}}]{Mouchet2001}
\bibinfo{author}{\bibfnamefont{A.}~\bibnamefont{Mouchet}},
  \bibinfo{author}{\bibfnamefont{C.}~\bibnamefont{Miniatura}},
  \bibinfo{author}{\bibfnamefont{R.}~\bibnamefont{Kaiser}},
  \bibinfo{author}{\bibfnamefont{B.}~\bibnamefont{Gr\'{e}maud}},
  \bibnamefont{and} \bibinfo{author}{\bibfnamefont{D.}~\bibnamefont{Delande}},
  \bibinfo{journal}{Phys. Rev. E} \textbf{\bibinfo{volume}{64}},
  \bibinfo{pages}{016221} (\bibinfo{year}{2001}).

\bibitem[{\citenamefont{Berry}(1978)}]{Berry78a}
\bibinfo{author}{\bibfnamefont{M.~V.} \bibnamefont{Berry}}, in
  \emph{\bibinfo{booktitle}{Topics in Nonlinear Dynamics --- A Tribute to Sir
  {E}dward {B}ullard}}, edited by \bibinfo{editor}{\bibnamefont{{S}iebe}}
  \bibnamefont{and} \bibinfo{editor}{\bibnamefont{{J}orna}}
  (\bibinfo{year}{1978}), vol.~\bibinfo{volume}{46}, pp.
  \bibinfo{pages}{16--120}, ISBN \bibinfo{isbn}{0-88318-145-2},
  \bibinfo{note}{reprinted in~\cite{Mackay/Meiss87a}}.

\bibitem[{\citenamefont{Arnold}(1979)}]{Arnold79}
\bibinfo{author}{\bibfnamefont{V.~I.} \bibnamefont{Arnold}},
  \emph{\bibinfo{title}{Mathematical methods of classical mechanics}}
  (\bibinfo{publisher}{Springer-Verlag, New York}, \bibinfo{year}{1979}).

\bibitem[{\citenamefont{Ashcroft and Mermin}(1976)}]{Ashcroft76}
\bibinfo{author}{\bibfnamefont{N.~W.} \bibnamefont{Ashcroft}} \bibnamefont{and}
  \bibinfo{author}{\bibfnamefont{N.~D.} \bibnamefont{Mermin}},
  \emph{\bibinfo{title}{Solid State Physics}} (\bibinfo{publisher}{Saunders
  College Philadelphia}, \bibinfo{year}{1976}).

\bibitem[{\citenamefont{Klauder and Skagerstam}(1985)}]{Klauder85}
\bibinfo{author}{\bibfnamefont{J.~R.} \bibnamefont{Klauder}} \bibnamefont{and}
  \bibinfo{author}{\bibfnamefont{B.-S.} \bibnamefont{Skagerstam}},
  \emph{\bibinfo{title}{Coherent States---Applications in Physics and
  Mathematical Physics}} (\bibinfo{publisher}{World Scientific},
  \bibinfo{address}{Singapore}, \bibinfo{year}{1985}), ISBN
  \bibinfo{isbn}{9971-966-52-0}.

\bibitem[{\citenamefont{Cohen-Tannoudji
  et~al.}(1977)\citenamefont{Cohen-Tannoudji, Diu, and Lalo\"{e}}}]{Cohen77}
\bibinfo{author}{\bibfnamefont{C.}~\bibnamefont{Cohen-Tannoudji}},
  \bibinfo{author}{\bibfnamefont{B.}~\bibnamefont{Diu}}, \bibnamefont{and}
  \bibinfo{author}{\bibfnamefont{F.}~\bibnamefont{Lalo\"{e}}},
  \emph{\bibinfo{title}{Quantum mechanics}}
  (\bibinfo{publisher}{Wiley-Interscience}, \bibinfo{year}{1977}).

\bibitem[{\citenamefont{Denschlag et~al.}(2002)\citenamefont{Denschlag,
  Simsarian, H\"{a}ffner, McKenzie, Browaeys, Cho, Helmerson, Rolston, and
  Phillips}}]{Denschlag2002}
\bibinfo{author}{\bibfnamefont{J.~H.} \bibnamefont{Denschlag}},
  \bibinfo{author}{\bibfnamefont{J.~E.} \bibnamefont{Simsarian}},
  \bibinfo{author}{\bibfnamefont{H.}~\bibnamefont{H\"{a}ffner}},
  \bibinfo{author}{\bibfnamefont{C.}~\bibnamefont{McKenzie}},
  \bibinfo{author}{\bibfnamefont{A.}~\bibnamefont{Browaeys}},
  \bibinfo{author}{\bibfnamefont{D.}~\bibnamefont{Cho}},
  \bibinfo{author}{\bibfnamefont{K.}~\bibnamefont{Helmerson}},
  \bibinfo{author}{\bibfnamefont{S.~L.} \bibnamefont{Rolston}},
  \bibnamefont{and} \bibinfo{author}{\bibfnamefont{W.~D.}
  \bibnamefont{Phillips}}, \bibinfo{journal}{J. Phys. B - At. Mol. Opt.}
  \textbf{\bibinfo{volume}{35}}, \bibinfo{pages}{3095} (\bibinfo{year}{2002}).

\bibitem[{\citenamefont{Mouchet and Delande}(2003)}]{Mouchet03a}
\bibinfo{author}{\bibfnamefont{A.}~\bibnamefont{Mouchet}} \bibnamefont{and}
  \bibinfo{author}{\bibfnamefont{D.}~\bibnamefont{Delande}},
  \bibinfo{journal}{Phys. Rev.E} \textbf{\bibinfo{volume}{67}},
  \bibinfo{pages}{046216} (\bibinfo{year}{2003}).

\bibitem[{\citenamefont{Brodier et~al.}(2001)\citenamefont{Brodier, Schlagheck,
  and Ullmo}}]{Brodier2001}
\bibinfo{author}{\bibfnamefont{O.}~\bibnamefont{Brodier}},
  \bibinfo{author}{\bibfnamefont{P.}~\bibnamefont{Schlagheck}},
  \bibnamefont{and} \bibinfo{author}{\bibfnamefont{D.}~\bibnamefont{Ullmo}},
  \bibinfo{journal}{Phys. Rev. Lett.} \textbf{\bibinfo{volume}{87}},
  \bibinfo{pages}{064101} (\bibinfo{year}{2001}).

\bibitem[{\citenamefont{Hensinger
  et~al.}(2001{\natexlab{b}})\citenamefont{Hensinger, Upcroft, Holmes,
  Heckenberg, Milburn, and Rubinsztein-Dunlop}}]{Hensinger2001a}
\bibinfo{author}{\bibfnamefont{W.~K.} \bibnamefont{Hensinger}},
  \bibinfo{author}{\bibfnamefont{B.}~\bibnamefont{Upcroft}},
  \bibinfo{author}{\bibfnamefont{C.~A.} \bibnamefont{Holmes}},
  \bibinfo{author}{\bibfnamefont{N.~R.} \bibnamefont{Heckenberg}},
  \bibinfo{author}{\bibfnamefont{G.~J.} \bibnamefont{Milburn}},
  \bibnamefont{and}
  \bibinfo{author}{\bibfnamefont{H.}~\bibnamefont{Rubinsztein-Dunlop}},
  \bibinfo{journal}{Phys. Rev. A} \textbf{\bibinfo{volume}{64}},
  \bibinfo{pages}{063408} (\bibinfo{year}{2001}{\natexlab{b}}).

\bibitem[{\citenamefont{Hensinger
  et~al.}(2001{\natexlab{c}})\citenamefont{Hensinger, Truscott, Upcroft, Hug,
  Wiseman, Heckenberg, and Rubinsztein-Dunlop}}]{Hensinger2001b}
\bibinfo{author}{\bibfnamefont{W.~K.} \bibnamefont{Hensinger}},
  \bibinfo{author}{\bibfnamefont{A.~G.} \bibnamefont{Truscott}},
  \bibinfo{author}{\bibfnamefont{B.}~\bibnamefont{Upcroft}},
  \bibinfo{author}{\bibfnamefont{M.}~\bibnamefont{Hug}},
  \bibinfo{author}{\bibfnamefont{H.~M.} \bibnamefont{Wiseman}},
  \bibinfo{author}{\bibfnamefont{N.~R.} \bibnamefont{Heckenberg}},
  \bibnamefont{and}
  \bibinfo{author}{\bibfnamefont{H.}~\bibnamefont{Rubinsztein-Dunlop}},
  \bibinfo{journal}{Phys. Rev. A} \textbf{\bibinfo{volume}{64}},
  \bibinfo{pages}{033407} (\bibinfo{year}{2001}{\natexlab{c}}).

\bibitem[{\citenamefont{Mackay and Meiss}(1987)}]{Mackay/Meiss87a}
\bibinfo{author}{\bibfnamefont{R.~S.} \bibnamefont{Mackay}} \bibnamefont{and}
  \bibinfo{author}{\bibfnamefont{J.~D.} \bibnamefont{Meiss}},
  \emph{\bibinfo{title}{Hamiltonian Dynamical Systems}}
  (\bibinfo{publisher}{{A}dam {H}ilger}, \bibinfo{address}{Bristol and
  Philadelphia}, \bibinfo{year}{1987}), ISBN \bibinfo{isbn}{0-85274-205-3}.

\end{thebibliography}
\end{document}